\documentclass[showpacs,preprintnumbers,two column]{revtex4-1}
\usepackage{amssymb}
\usepackage{amsmath}
\usepackage{graphicx}
\usepackage{dcolumn}
\usepackage{bm}

\begin{document}

\title{Generalized squeezing rotating-wave approximation to the isotropic
and anisotropic Rabi model in the ultrastrong-coupling regime}
\author{Yu-Yu Zhang}
\address{Department of Physics, Chongqing University, Chongqing
401330, People's Republic of China
}

\date{\today }

\begin{abstract}
Generalized squeezing rotating-wave approximation (GSRWA) is proposed by
employing both the displacement and the squeezing transformations. A solvable
Hamiltonian is reformulated in the same form as the ordinary RWA ones. For a
qubit coupled to oscillators experiment, a well-defined Schr\"{o}%
dinger-cat-like entangled state is given by the displaced-squeezed
oscillator state instead of the original displaced state. For the isotropic
Rabi case, the mean photon number and the ground-state energy are expressed
analytically with additional squeezing terms, exhibiting a substantial
improvement of the GSRWA. And the ground-state energy in the anisotropic Rabi model
confirms the effectiveness of the GSRWA. Due to
the squeezing effect, the GSRWA improves the previous methods only with the
displacement transformation in a wide range of coupling strengths even for
large atom frequency.
\end{abstract}

\maketitle

\section{Introduction}

The quantum Rabi model~\cite{Rabi} describes the coupling of a two-level
atom to a bosonic field in quantum optics and has been attracting a
remarkable amount of interest since it is now the focus of many recent
applications in cavity~\cite{baumann,nagy,Dimer} and circuit~\cite%
{Wallraff,Niemczyk,pfd,fedorov} quantum electrodynamics (QED). In early work
on cavity QED, coupling strength achieved was much smaller than the
frequency of the field. In this way, the Rabi model could be drastically
simplified to the Jaynes-Cumming model through the rotating-wave
approximation (RWA)~\cite{jaynes}. In recent circuit QED setups, where
artificial superconducting two-level atoms are coupled to on-chip cavities,
it is evident for the breakdown of the RWA in the ultrastrong-coupling
regime~\cite{Niemczyk,pfd} and the counter-rotating-wave (CRW) terms are
expected to take effect. Interestingly, different coupling strengths of the
rotating-wave (RW) and CRW interactions can be tuned independently by
applied electric and magnetic fields in recent cavity and circuit QED setups~%
\cite{erlingsson,ye}. Much attention has been paid over a generalization of the
Rabi model called the anisotropic Rabi model~\cite{xie,judd,grimsno} with
different coupling strengths for the RW and CRW interactions.

The isotropic and anisotropic Rabi model lacks a full closed-form
solution due to the inclusion of the CRW terms. Despite the fact that the exact
solution has been given by a Bargmann space technique~\cite%
{xie,judd,braak1,tomka} and an extended coherent state method~\cite{chen},
respectively, where a numerical search for the zeros of complicated
transcendental functions is needed, an efficient, easy-to-implement
theoretical treatment remains elusive. There have been numerous theoretical
studies on the Rabi model finding approximated solutions~\cite%
{irish,zhang,agarwal,Ashhab,yu,victor}. A generalized RWA (GRWA) with fixed
displacement~\cite{irish,zhang} fails to describe the ground state
especially for a large atom frequency, since it is equivalent to the
adiabatic approximation in the ground state~\cite{agarwal,Ashhab}. As an
improvement, a generalized variational method (GVM) frees the displacement,
but breaks down for the ultra-strong coupling, where the coupling strength is
$0.5$ of the oscillator frequency~\cite%
{yu,zheng}. And an analytical method with the displacement transformation
was employed to study the anisotropic Rabi model~\cite{zhu}, which was
restricted to the relatively weak coupling for small atom frequency. As
previously studied, the oscillator state was considered as a displaced state,
and the squeezing effect of the oscillator part is underestimated. Moreover,
it gives rise to interesting phenomena when the atomic frequency in the
unit of the cavity frequency increases~\cite{ying,plenio}. So it is highly
desirable to study the solution in a wide range of coupling strengths
especially for a large atom frequency.

Motivated by those developments, we propose a generalized squeezing RWA
(GSRWA) by adding a squeezing transformation to the original displacement
transformation. A reformulated Hamiltonian with the same form as the
ordinary RWA one is evaluated. We investigate the role of these two variable displacement and
squeezing in lowering the ground-state energy. In particular, an analytical
solution to the isotropic Rabi model is obtained approximately, giving the
mean photon number and the ground-state energy with the additional squeezing
terms. We also obtain reasonable accurate ground-state
energy after the level-crossing point in the anisotropic Rabi model. Furthermore,
validity of our approach is discussed by comparing with the GVM and GRWA as well
as numerical simulation.

The paper is outlined as follows. In Sec.~II, we derive the effective
Hamiltonian through both the displacement and the squeezing transformations. In
Sec.~III, an analytical solution to the isotropic Rabi model is derived
approximately. The entanglement, the ground-state energy and the mean photon
number are given analytically. In the anisotropic case, the ground-state
energy is calculated by the GSRWA. Finally, a brief summary is given in
Sec.~IV.

\section{Isotropic and anisotropic Rabi model}

The anisotropic Rabi model describes a two-level atom coupled to a bosonic
field with different coupling strengths of the RW and CRW interactions,
yielding the Hamiltonian as follows:
\begin{equation}
H=\frac{\Delta }{2}\sigma _{z}+\omega a^{\dagger }a+g_{1}\left( a^{\dagger
}\sigma _{-}+a\sigma _{+}\right) +g_{2}\left( a^{\dagger }\sigma
_{+}+a\sigma _{-}\right) ,  \label{Hamiltonian}
\end{equation}%
where $\sigma _{x}$ and $\sigma _{z}$ are Pauli matrices for the two-level
atom with transition frequency $\Delta $, $a^{\dagger }$ $\left( a\right) $
is the creation (annihilation) operator of the single-mode bosonic field
with frequency $\omega $, $g_{1}\ $and $g_{2}\ $ are the coupling strength
of the rotating terms $a^{\dagger }\sigma _{-}+a\sigma _{+}$ and
counter-rotating terms $a^{\dagger }\sigma _{+}+a\sigma _{-}$, respectively.
It reduces to the isotropic Rabi model under the assumption $g_{1}=g_{2}$.
In the paper, we take the unit of $\omega =1$.

For convenience, the Hamiltonian is written into
\begin{equation}
H=\frac{\Delta }{2}\sigma _{z}+\omega a^{\dagger }a+\alpha (a^{\dagger
}+a)\sigma _{x}+i\sigma _{y}\gamma \left( a^{\dagger }-a\right) ,
\label{hamiltonian}
\end{equation}%
where$\;\alpha =\left( g_{1}+g_{2}\right) /2$ and$\;\gamma =\left(
g_{2}-g_{1}\right) /2$. Performing a unitary transformation $U=\exp \left[
\beta \sigma _{x}\left( a^{+}-a\right) \right] $ with a dimensionless
variable displacement $\beta $, one obtains $H_{1}=UHU^{\dagger }$ with
\begin{eqnarray}
H_{1} &=&\omega a^{\dagger }a+\omega \beta ^{2}-2\beta \omega \alpha
+(\alpha -\omega \beta )(a^{\dagger }+a)\sigma _{x}  \notag  \label{coherent}
\\
&&+\frac{\Delta }{2}\{\sigma _{z}\cosh [2\beta \left( a^{\dagger }-a\right)
]-i\sigma _{y}\sinh [2\beta \left( a^{\dagger }-a\right) ]\}  \notag \\
&&+\gamma \left( a^{\dagger }-a\right) \{-\sigma _{z}\sinh [2\beta \left(
a^{\dagger }-a\right) ]  \notag \\
&&+i\sigma _{y}\cosh [2\beta \left( a^{\dagger }-a\right) ]\}.
\end{eqnarray}%
Moreover, we employ the bosonic squeezed operator with a variational
squeezing parameter $\lambda $
\begin{equation}
V=e^{\lambda (a^{2}-a^{+2})},
\end{equation}%
acting on $a^{\dagger }$ and $a$, which yields $VaV^{\dagger }=a^{\dagger
}\sinh 2\lambda +a\cosh 2\lambda $ and $Va^{\dagger }V^{\dagger }=a\sinh
2\lambda +a^{\dagger }\cosh 2\lambda $. Neglecting the two-photon terms of $%
a^{\dagger 2}$ and $a^{2}$ , the Hamiltonian becomes $H^{\prime
}=VH_{1}V^{\dagger }=H_{0}^{\prime }+H_{1}^{\prime }$, consisting of
\begin{eqnarray}
H_{0}^{\prime } &=&\eta _{0}+\eta _{1}a^{\dagger }a+\sigma _{z}\{\frac{%
\Delta }{2}\cosh [2\beta \eta \left( a^{\dagger }-a\right) ]  \notag
\label{squeezed} \\
&&-\gamma \left( a^{\dagger }-a\right) \eta \sinh [2\beta \eta \left(
a^{\dagger }-a\right) ]\}, \\
H_{1}^{\prime } &=&\eta _{2}\sigma _{x}(\alpha -\omega \beta )(a^{\dagger
}+a)+i\sigma _{y}\{-\frac{\Delta }{2}\sinh [2\beta \eta \left( a^{\dagger
}-a\right) ]  \notag \\
&&+\gamma \left( a^{\dagger }-a\right) \eta \cosh [2\beta \eta \left(
a^{\dagger }-a\right) ]\},  \label{h1}
\end{eqnarray}%
where $\eta _{0}=\omega \sinh ^{2}2\lambda +\omega \beta ^{2}-2\beta \alpha $%
, $\eta _{1}=\omega (\cosh ^{2}2\lambda +\sinh ^{2}2\lambda )$, $\eta
_{2}=\cosh 2\lambda +\sinh 2\lambda $, and $\eta =\omega (\cosh 2\lambda
-\sinh 2\lambda )$. In this paper, we employ both the displacement and
squeezing transformation to tackle the isotropic and anisotropic Rabi model.
In contrast to the previous methods with only the displacement
transformation, such as the GVM or GRWA, our method captures more essential
physics due to the inclusion of the bosonic squeezing transformation.

First, we expand the even and odd functions $\cosh (y)$ and $\sinh (y)$ by
keeping the leading terms. When $\cosh [2\beta \eta (a^{\dagger }-a)]$ is
expanded as $1+\frac{1}{2!}[2\beta \eta \left( a^{\dagger }-a\right) ]^{2}+%
\frac{1}{4!}[2\beta \eta \left( a^{\dagger }-a\right) ]^{4}+...$, it is
performed by keeping the terms containing the number operator $a^{\dagger
}a=n$ as
\begin{equation}
\cosh [2\beta \eta \left( a^{\dagger }-a\right) ]=G(a^{\dagger }a)+O(\beta
^{2}\eta ^{2}),
\end{equation}%
where the coefficient $G(a^{\dagger }a)$ that depends on the oscillator
number operator $a^{\dagger }a$ and the dimensionless parameters $\beta$ and
$\eta$ can be expressed in the oscillator basis $|n\rangle $ as
\begin{equation}
G_{n,n}=\langle n|\cosh [2\beta \eta (a^{\dagger }-a)]|n\rangle =\chi
L_{n}(4\beta ^{2}\eta ^{2}).
\end{equation}
Here $\chi =\exp (-2\beta ^{2}\eta ^{2})$, and $L_{n}^{m-n}$ are the
Laguerre polynomials. $O(\beta ^{2}\eta ^{2})$ is accounted for higher-order
(multi-photon) processes, which is neglected within this approximation.
Similarly, the odd function $\sinh [2\beta \eta \left( a^{\dagger }-a\right)
$ can be expanded by keeping the one-excitation terms as%
\begin{equation}
\sinh [2\beta \eta \left( a^{\dagger }-a\right) ]=R(a^{\dagger
}a)a^{+}-aR(a^{\dagger }a)+O(\beta ^{3}\eta ^{3}).
\end{equation}%
Since the terms $aR\left( a^{\dagger }a\right) $ and $R\left( a^{\dagger
}a\right) a^{\dagger }$ involve creating and eliminating a single photon of
the oscillator, it is reasonable to set $\left\langle n+1\right\vert
R_{n+1,n}a^{\dagger }\left\vert n\right\rangle =\left\langle n+1\right\vert
R\left( a^{\dagger }a\right) a^{\dagger }\left\vert n\right\rangle $ and $%
\left\langle n\right\vert aR_{n,n+1}\left\vert n+1\right\rangle
=\left\langle n\right\vert aR\left( a^{\dagger }a\right) \left\vert
n+1\right\rangle $ by
\begin{eqnarray}
R_{n+1,n} &=&\frac{1}{\sqrt{n+1}}\langle n+1|\sinh [2\beta \eta \left(
a^{\dagger }-a\right) ]|n\rangle  \notag \\
&=&\frac{2\beta \eta \chi }{n+1}L_{n}^{1}(4\beta ^{2}\eta ^{2}).
\end{eqnarray}%
Following the similar approximation, the other operators in Eqs.(~\ref%
{squeezed}) and (~\ref{h1}) have the following leading terms:%
\begin{eqnarray}
\left( a^{\dagger }-a\right) \sinh [2\beta \eta \left( a^{\dagger }-a\right)
] &=&F(a^{\dagger }a)+O(\beta ^{2}\eta ^{2}), \\
\left( a^{\dagger }-a\right) \cosh [2\beta \eta \left( a^{\dagger }-a\right)
] &=&T(a^{\dagger }a)a^{+}-aT(a^{\dagger }a)  \notag \\
&&+O(\beta ^{3}\eta ^{3}),
\end{eqnarray}%
where the coefficients $F(a^{\dagger }a)$ and $T(a^{\dagger }a)$ are derived
in the oscillator basis $|n\rangle$ as, respectively,
\begin{eqnarray}
F_{n,n} &=&\langle n|\left( a^{\dagger }-a\right) \sinh [2\beta \eta \left(
a^{\dagger }-a\right) ]|n\rangle  \notag \\
&=&-\sqrt{n}R_{n,n-1}-\sqrt{n+1}R_{n+1,n}, \\
T_{n+1,n} &=&\frac{1}{\sqrt{n+1}}\langle n+1|\left( a^{\dagger }-a\right)
\cosh [2\beta \eta \left( a^{\dagger }-a\right) ]|n\rangle  \notag \\
&=&G_{n,n}-\sqrt{\frac{n+2}{n+1}}G_{n+2,n},
\end{eqnarray}%
with $G_{n+2,n}=4\beta ^{2}\eta ^{2}\chi L_{n}^{2}(4\beta ^{2}\eta ^{2})/%
\sqrt{(n+1)(n+2)}$.

After such a simplification, we have the reformulated Hamiltonian
\begin{eqnarray}
H_{\mathtt{GSRWA}} &=&\eta _{0}+\eta _{1}a^{\dagger }a+\sigma _{z}[\frac{%
\Delta }{2}G(a^{\dagger }a)-\gamma \eta F(a^{\dagger }a)]  \notag
\label{effective ham} \\
&&+[\eta _{2}(\alpha -\omega \beta )+\frac{\Delta }{2}R(a^{\dagger
}a)-\gamma \eta T(a^{\dagger }a)]a^{\dagger }\sigma _{-}+h.c.  \notag \\
&&
\end{eqnarray}%
One can easily diagonalize $H_{\mathtt{GSRWA}}$ in the basis of $%
|n,+z\rangle $ and $|n+1,-z\rangle $,
\begin{widetext}
\begin{equation}
H_{\mathtt{GSRWA}}=\left(
\begin{array}{cc}
\eta _{0}+n\eta _{1}+f(n) & [\eta _{2}(\alpha -\omega \beta )+P_{n+1,n}]%
\sqrt{n+1} \\
\lbrack \eta _{2}(\alpha -\omega \beta )+P_{n+1,n}]\sqrt{n+1} & \eta
_{0}+(n+1)\eta _{1}-f(n+1)%
\end{array}%
\right) ,
\end{equation}%
\end{widetext}
where $f(n)=\frac{\Delta }{2}G_{n,n}-\gamma \eta F_{n,n}$ and $P_{n+1,n}=%
\frac{\Delta }{2}R_{n+1,n}-\gamma \eta T_{n+1,n}$. Eigenstates and
eigenvalues can be given easily, which depends on the variable displacement
and squeezing.

Here, the terms retained in the reformulated Hamiltonian $H_{\mathtt{GSRWA}}$
correspond to the energy-conserving terms, just as in the
standard RWA. The dominated effect of the original CRW
interaction is considered here, since the atom frequency and the coupling
strength of the RW terms are renormalized induced by the CRW term. The
present GSRWA approach essentially extends the basic idea of the GRWA~\cite%
{irish,zhang} by adding a squeezing to its original displacement
transformation for the Rabi model. The merit of
the GSRWA is based on its connection to the additional squeezing and the
mathematical simplicity of the usual RWA, predicting a non-trivial
improvement over the previous GVM or GRWA, especially in the ultra-strong coupling regime.

In previous studies, the GRWA method with a fixed displacement works well in
a wide range of coupling regime~\cite{irish,zhang}. But it fails to give
correct ground-state energy for the ultra-strong coupling strength close to the oscillator frequency especially for
large atom frequency $\Delta \geq\omega $. In this paper one has the
displacement $\beta=\alpha/\omega$ for the GRWA. On the other hand, the GVM
frees the displacement and improves the GRWA in the relative weak coupling
regime, $g_1/\omega\sim0.5$, for small atom frequency~\cite{yu}. But it has been excessively simplified in the analytical
treatment, resulting in an incorrect ground state in intermediate and strong
coupling regimes. It is worthwhile to study the ground state in the
isotropic and anisotropic case by the GSRWA which, in return, works well
with the squeezing even for a large atom
frequency in the ultra-strong coupling regime, $g_1/\omega\sim1$.

\section{Ground state and ground-state energy}

As an improvement, the GSRWA is presented by adding the squeezing
transformation to the original displacement transformation. A lowering of
the ground-state energy is achieved for an optimum value of the displacement
$\beta$ and squeezing $\lambda$. We focus on the ground-state energy and
mean photon number for the large atom frequency to show the validity of the
GSRWA.

\textsl{Analytical solution for the isotropic Rabi model ($\gamma =0$):} It
is interesting to discuss the improvement of the analytical expression of
the ground state for the isotropic Rabi model, which has the same
coupling strength for the RW and CRW terms. In this case $\gamma=0$, it
allows us to give an analytical solutions to the optimal $\beta$ and $%
\lambda $.

From the reformulated Hamiltonian $H_{\mathtt{GSRWA}}$ (~\ref{effective ham}%
), the ground state is obviously
\begin{equation}
|G\rangle =|0\rangle |-z\rangle ,  \label{gs}
\end{equation}%
where $|0\rangle $ is the vacuum state of the oscillator, and $|-z\rangle $
is the eigenstate of $\sigma _{z}$ with eigenvalue $-1$. Note that the
displacement transformations $U=\exp \left[ \beta \sigma _{x}\left(
a^{+}-a\right) \right] $ and the squeezing transformation $V=e^{\lambda
(a^{2}-a^{+2})}$ can recast $|G\rangle $ to $|\tilde{G}\rangle =U^{\dagger
}V^{\dagger }|G\rangle $ as
\begin{eqnarray}  \label{gs1}
|\tilde{G}\rangle &=&\frac{1}{\sqrt{2}}[e^{-\beta \left( a^{+}-a\right)
}|0\rangle _{s}|+x\rangle +e^{\beta \left( a^{+}-a\right) }|0\rangle
_{s}|-x\rangle ],  \notag \\
&&
\end{eqnarray}%
where $|\pm x\rangle =\frac{1}{\sqrt{2}}(\mp |+z\rangle +|-z\rangle )$ are
the eigenstates of $\sigma _{x}$ with eigenvalues $\pm 1$. And $|0\rangle
_{s}$ denotes the squeezed vacuum state of the oscillator as $V^{\dagger
}|0\rangle $. We observe that the oscillator state is the displaced-squeezed
state $e^{\pm \beta \left( a^{+}-a\right) }|0\rangle _{s}$, which is
different from the original displaced oscillator state $e^{\pm \beta \left(
a^{+}-a\right) }|0\rangle $ in the GRWA or GVM. This is an advantage to employ
the GSRWA as an efficient approach to give a more accurate solution.

For the isotropic Rabi model with $\gamma =0$, the ground-state energy is
evaluated as
\begin{equation}
E_{gs}^{\mathtt{GSRWA}}=\omega \sinh ^{2}2\lambda +\omega \beta ^{2}-2\beta
\alpha -\frac{\Delta }{2}e^{-2\beta ^{2}\exp (-4\lambda )}.  \label{gsGRWA}
\end{equation}%
It is interesting to note that $E_{gs}^{\mathtt{GSRWA}}$ includes the
additional squeezing terms in the first and the last terms. And the
last three terms with $\lambda =0$ basically give the results in the GVM~%
\cite{yu} as $E_{gs}^{\mathtt{GVM}}=\omega \beta ^{2}-2\beta \alpha -\frac{%
\Delta }{2}e^{-2\beta ^{2}}$, which has the optimal value when $\beta \simeq
\frac{\alpha }{\omega +\Delta e^{-2\alpha ^{2}/(\omega +\Delta )^{2}}}$.

\begin{figure}[tbp]
\includegraphics[scale=0.5]{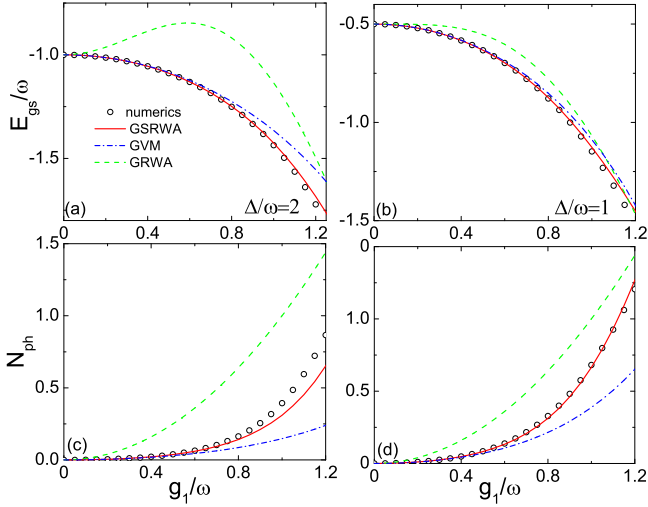}
\caption{(Color online) (a) and (b) The ground-state energy $E_{gs}/\protect\omega$
and (c) and (d) the corresponding mean photon number $N_{ph}$ (c-d) obtained
analytically in the isotropic case as a function of $g_1/\protect\omega$ for
different detuning cases obtained by the GSRWA (solid lines), numerical
simulation (circles), GVM (dashed-dotted lines), and GRWA (dashed lines) for
the isotropic Rabi model.}
\label{analytical}
\end{figure}
Within the GSRWA, the optimal variables $\beta $ and $\lambda $ can be
derived from minimizing the ground-state energy according to $\partial
E_{gs}^{\mathtt{GSRWA}}/\partial \beta =0$ and $\partial E_{gs}^{\mathtt{%
GSRWA}}/\partial \lambda =0$ by
\begin{eqnarray}
0 =\beta \Delta e^{-4\lambda }e^{-2\beta ^{2}\exp (-4\lambda )}+\omega \beta
-\alpha ,  \label{mini1}
\end{eqnarray}
\begin{eqnarray}
0 =-4\beta ^{2}\Delta e^{-4\lambda }e^{-2\beta ^{2}\exp (-4\lambda )}+\omega
(e^{4\lambda }-e^{-4\lambda }).  \label{mini2}
\end{eqnarray}
When the dimensionless parameters $\lambda$ and $\beta$ are less than $1$,
the displacement $\beta $ is given approximately by
\begin{equation}  \label{beta}
\beta \simeq \frac{\alpha }{\omega +\Delta e^{-4\lambda }e^{-2\alpha
^{2}/(\omega +\Delta )^{2}\exp (-4\lambda )}},
\end{equation}%
and the squeezing parameter is
\begin{equation}  \label{lambda}
\lambda \simeq \frac{\Delta }{2\omega}\frac{\alpha ^{2}}{(\omega +\Delta
)^{2}}.
\end{equation}%
Interestingly, $\lambda $ depends on the atom frequency $\Delta $, resulting
in an important squeezing effect in lowering the ground-state energy as $%
\Delta $ increases. And $\beta $ is amended by adding the
squeezing parameter as $\exp (-4\lambda )$ by comparing with $\beta$ in the
GVM. With the increase in atom frequency $\Delta$ and the effective coupling
strength $\alpha$, both the displacement and the squeezing play an important
role in lowering the ground-state energy.

In particular, in the current experimental setups of the ultra-strong
coupling, $g_1/\omega,g_2/\omega\simeq0.1$, the optimal displacement $\beta$
is simplified as $\alpha/(\omega +\Delta)$, and one can obtain an explicit
ground-state energy
\begin{eqnarray}
E_{gs}^{\mathtt{GSRWA}}&&\simeq \omega \sinh ^{2}[\frac{\Delta\alpha ^{2}}{%
\omega(\omega +\Delta )^{2}}]-\frac{\alpha ^{2}(\omega +2\Delta )}{(\omega +\Delta
)^{2}}  \notag \\
&&-\frac{\Delta }{2}e^{-\frac{2\alpha ^{2}}{(\omega +\Delta )^{2}} \exp [-%
\frac{2\Delta \alpha ^{2}}{(\omega +\Delta )^{2}} ]},
\end{eqnarray}
which exhibits an improvement over the one in the GVM $E_{gs}^{\mathtt{GVM}%
}=-\frac{\alpha ^{2}(\omega +2\Delta )}{(\omega +\Delta )^{2}} -\frac{\Delta%
}{2}\mathtt{exp}[-\frac{2\alpha ^{2}}{(\omega +\Delta )^{2}}]$~\cite{yu}.

On the other hand, in the deep-strong coupling limit, $g_1/\omega\gg1$, we
derive approximately $\beta \simeq \alpha /\omega $ and $\lambda \simeq 0$
from the optimal Eqs. (~\ref{mini1}) and (~\ref{mini2}). Obviously, the
squeezing effect disappears. The ground-state energy becomes
\begin{equation}
E_{gs}^{\mathtt{GSRWA}}\simeq -\frac{\alpha ^{2}}{\omega }-\frac{\Delta }{2}%
e^{-\frac{2\alpha^2}{\omega^2}},
\end{equation}%
which is the same as that obtained by the GRWA with $\beta =\alpha /\omega $.

Figures ~\ref{analytical} (a) and (b) show the ground-state energy in the GSRWA
obtained analytically by the displacement $\beta$ (~\ref{beta}) and the
squeezing $\lambda$ (~\ref{lambda}) for large atom frequency. For weak
coupling strength $g_1/\omega$, contribution from the coupling is small and
the ground-state energies obtained by the four methods are equal. As $g_1$
increases, the oscillator state becomes the displaced state due to the
coupling of the atom and the original oscillator. The variable displacement
in the GSRWA or GVM becomes to play a role in lowering the ground-state
energy, exhibiting lower energy than the GRWA results. As $g_1$ enters
the ultra-strong coupling regime, $g_1/\omega\sim 0.5$, the squeezing effect
becomes appreciable and the oscillator state should be a displaced-squeezed
state instead of a displaced state. As a result, the GSRWA results agree
well with the numerical ones ranging from weak to ultra-strong coupling regimes.
However, the results from the GVM become worse for the ultra-strong coupling
strength $g_1/\omega>0.8$, especially for the positive detuning $\Delta/\omega=2$ in Fig.~\ref%
{analytical} (a). The deviations of the GVM or GRWA become more
noticeable as $\Delta/\omega$ increases. The reason for the failure is that
the squeezing effect is underestimated in the GVM and GRWA with only
the displacement transformation. Therefore, the squeezing effect becomes important
as the coupling strength increases up to $g_1/\omega\sim1$, especially for large atom frequency.

\begin{figure}[tbp]
\includegraphics[scale=0.7]{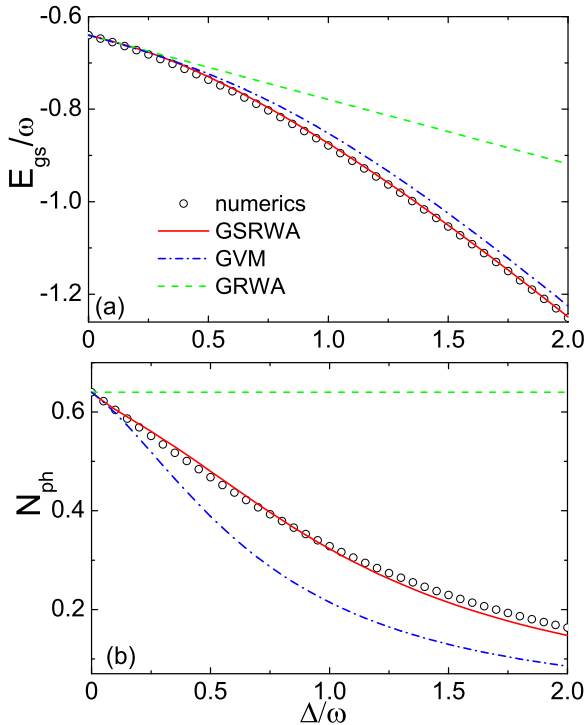}
\caption{(Color online) (a)The ground-state energy $E_{gs}/\protect\omega$
and (b) the mean photon number $N_{\mathtt{ph}}$ in the isotropic case as a
function of $\Delta/\protect\omega$ for the coupling strength $g_1/\protect%
\omega=0.8$ by means of the GSRWA (solid lines), numerical simulation
(circles), GVM (dash-dotted lines) and GRWA (dashed lines).}
\label{delta}
\end{figure}

To show the squeezing effect dependent on the detunings $\Delta/\omega$,
Fig.~\ref{delta}(a) shows the ground-state energy as a function of $%
\Delta/\omega$ for the ultra-strong coupling strength $g_1/\omega=0.8$. It shows
an accuracy of the ground-state energy in the GSRWA for arbitrary detunings due to
the squeezing effect in Eq.(~\ref{lambda}) dependent on $%
\Delta/\omega$.
However, there is a distinguished deviation of the GVM and
GRWA results from negative detuning $\Delta/\omega<1$ to positive detuning $%
\Delta/\omega>1$. It reveals that the squeezing plays an important effect in
lowering the ground-state energy as $\Delta/\omega$ increases.

We now discuss the mean photons number $N_{ph}=\langle a^{+}a\rangle $ in
the ground state $|\tilde{G}\rangle$, which can be derived by
\begin{equation}
N_{ph}^{\mathtt{GSRWA}} =\omega\sinh ^{2}2\lambda +\omega \beta ^{2}.
\end{equation}%
The second term denotes the mean photon number in the GVM, and the first
squeezing term shows an improvement of $N_{ph}$ by the GSRWA. One can obtain
the mean photon number in the GVM ~\cite{yu} as
\begin{equation}
N_{ph}^{\mathtt{GVM}} =\omega\beta^2\simeq\frac{\omega\alpha^2 }{[\omega
+\Delta e^{-2\alpha ^{2}/(\omega +\Delta )^{2}}]^2}.
\end{equation}
For the GRWA with $\beta=\alpha/\omega$, the mean photon number simply is
given by $N_{ph}^{\mathtt{GRWA}}=\omega\beta^2\simeq\alpha^2/\omega$, which
is independent of $\Delta$. In terms of the analytical solution of the
displacement $\beta$ (~\ref{beta}) and squeezing $\lambda$ (~\ref{lambda}),
the mean photon number under the GSRWA can explicitly be given by
\begin{eqnarray}
N_{ph}^{\mathtt{GSRWA}} &&\simeq \omega \sinh ^{2}[\frac{\Delta \alpha ^{2}}{%
\omega(\omega +\Delta )^{2}}]  \notag \\
&&+\frac{\omega\alpha^2 }{[\omega +\Delta e^{-4\lambda }e^{-2\alpha
^{2}/(\omega +\Delta )^{2}\exp (-4\lambda )}]^2}.
\end{eqnarray}%
It is observed that there is an oversimplified analytical treatment for the
mean photon number in the GVM and GRWA, but the GSRWA with additional
squeezing terms provides an efficient, yet accurate analytical expression.

The mean photon number is plotted as a function of the coupling strength $%
g_1/\omega$ for the positive detuning and resonance cases in Figs.~\ref{analytical}
(c)(d). The GSRWA agrees well with the numerical ones from weak to ultra-strong
coupling regimes. However, the GVM and GRWA results are qualitatively incorrect
as the coupling strength increases.

Behavior of the mean photon number $N_{ph}$ dependent on $\Delta/\omega$ is
plotted in Fig.~\ref{delta} (b) for the ultra-strong coupling strength $%
g_1/\omega=0.8$. The results derived from the GSRWA are consistent with those
of the numerical simulation from negative to positive detunings. For the
large negative detuning $\Delta/\omega\ll1$, the GVM results agree well with
the numerical ones. However, as $\Delta/\omega$ increases, the GVM breaks
down and the deviations becomes larger. And the GRWA keeps an incorrect
constant in the whole detuning case, since the analytical $N_{ph}^{\mathtt{%
GRWA}}$ is independent of $\Delta$. It is concluded that the analytical
results of the GSRWA with the squeezing is more accurate than those of the
previous GVM or GRWA for arbitrary detuning even for the ultra-strong coupling
strength $g_1/\omega\sim1$.

\begin{figure}[tbp]
\includegraphics[scale=0.75]{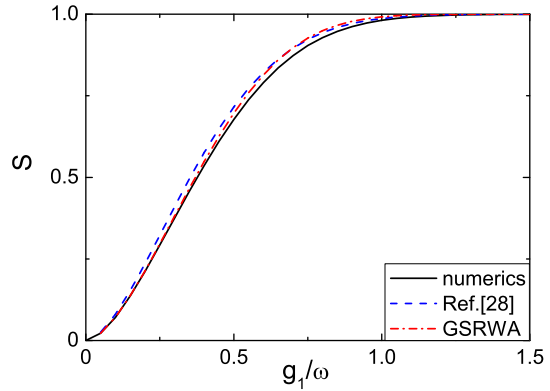}\label{entropy1}
\caption{(Color online) The qubit-oscillator entanglement $S$ obtained by
the GSRWA (dashed line) with the displacement in Eq.~(\protect\ref{beta})
and squeezing in Eq.~(\protect\ref{lambda}) using the experiment parameters
in Ref.~\protect\cite{yoshihara}. The entanglement obtained by numerical
simulation (solid line), and results in Ref.~\protect\cite{yoshihara}
(dashed-dotted line) are plotted for comparison.}
\end{figure}

\begin{center}
\begin{figure*}[tbp]
\includegraphics[scale=0.7]{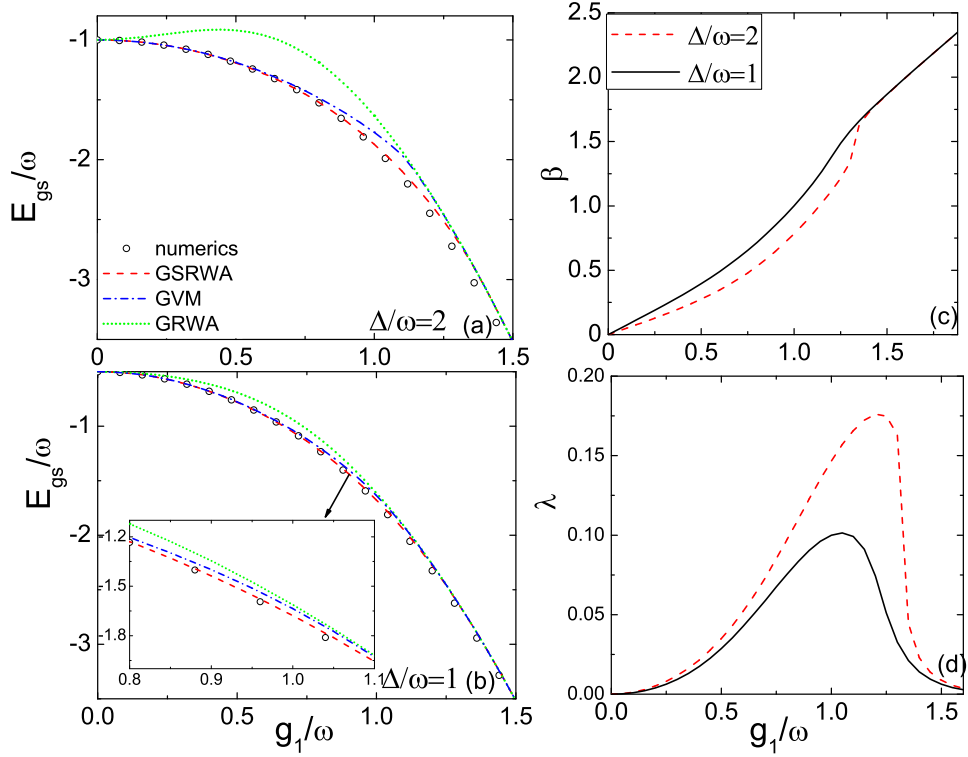}
\caption{(Color online) (a) and (b) The ground-state energy $E_{gs}/\protect\omega$
(a-b) in the anisotropic Rabi model with $\protect\gamma>0$ as a function of
$g_1/\protect\omega$ for different detuning cases by means of the GSRWA
(dashed lines), numerical simulation (circles), GVM (dashed-dotted lines) and
GRWA (dotted lines) with the coupling strength $g_2=1.5g_1$. The
corresponding (c) variable displacement $\protect\beta$ and (d) squeezing $%
\protect\lambda$ (d) as a function of $g_1/\protect\omega$ obtained by the
GSRWA for different detunings $\Delta/\protect\omega=2$ (dashed line) and $%
\Delta/\protect\omega=1$ (solid line).}
\label{large g2}
\end{figure*}
\end{center}

Due to the recent advances in circuit QED setups, the ground state can be
described by the Schr\"{o}dinger-cat-like entangled state between the qubit
and the oscillator beyond the ultra-strong coupling, where the oscillator state
was described as the displaced state $\mathtt{exp}[\pm g_1/\omega
(a^{\dagger}-a)]$~\cite{yoshihara}. In our method, the ground state $|\tilde{%
G}\rangle $ ~(\ref{gs1}) is improved by the displaced-squeezed oscillator
state, which is expected to be a well-defined Schr\"{o}dinger-cat-like
entangled state. The progress is made by using the entangled state $|\tilde{G%
}\rangle $ to estimate the qubit-oscillator entanglement $S$. Considering
the expression $V^{+}e^{\beta \left( a^{+}-a\right) }V=e^{\beta ^{\prime
}(a^{+}-a)}$ with $\beta ^{\prime }=\beta \eta_2$, the overlap of two
displaced-squeezed oscillator states is obtained as $_{s}\langle 0|e^{\beta
\left( a-a^{+}\right) }e^{-\beta \left( a^{+}-a\right) }|0\rangle
_{s}=e^{-2\beta ^{\prime 2}}$. It leads to the qubit's reduced density
matrix $\rho _{a}$ by tracing out the oscillator degrees of freedom,
\begin{equation*}
\rho _{a}=\frac{1}{2}\left(
\begin{array}{cc}
1 & e^{-2\beta ^{\prime 2}} \\
e^{-2\beta ^{\prime 2}} & 1%
\end{array}%
\right) ,
\end{equation*}%
where the eigenvalues of $\rho _{a}$ are $p_{\pm }$ $=(1\pm e^{-2\beta
^{\prime 2}})/2$. The entanglement can be measured by the von Neumann
entropy~\cite{Popescu,Wang} of the qubit:
\begin{equation}  \label{entropy}
S=-p_{+}\log _{2}p_{+}-p_{-}\log _{2}p_{-}.
\end{equation}

The entanglement is modified by $\beta ^{\prime }$ instead of the
displacement $\beta=g_1/\omega $ in previous results~\cite{yoshihara}. Fig.3
shows the qubit-oscillator entanglement $S$ using the analytical displacement in Eq.~(%
\ref{beta}) and the squeezing in Eq.~(\ref{lambda}), with fitted parameters
of the experimental results $\omega/2\pi=5.711~\mathrm{GHz}$ and $%
\Delta=0.441~\mathrm{GHz}$. Besides the validity of the GSRWA as well as of
Ref.~\cite{yoshihara} in the deep-strong coupling regime, $g_1/\omega>1$, a
noticeable improvement by the GSRWA is observed in the ultra-strong coupling regimes. This demonstrates the great potential of our
GSRWA approach to be applied in future experiments for the ultra-strong
coupling strength by the displaced-squeezed oscillator state instead of the original displaced state.

\textsl{Dominated coupling of the CRW interaction ($\gamma >0$):} For the
anisotropic Rabi model, the different couplings allow us to vary the CRW
interaction independently and explore the effects of it on some quantum
properties. We first study the ground-state energy in the case of the
dominated coupling of the CRW term, $g_{2}\geq g_{1}$ with $\gamma >0$.

Averaging over the ground state $|G\rangle $ (~\ref{gs}) the ground-state
energy is obtained as
\begin{eqnarray}
E_{gs}^{\mathtt{GSRWA}}&&=\omega \sinh ^{2}2\lambda +\omega \beta
^{2}-2\beta \alpha  \notag \\
&&-(\frac{\Delta }{2}+2\gamma \beta e^{-4\lambda })e^{-2\beta ^{2}\exp
(-4\lambda )}.  \label{gs energy}
\end{eqnarray}%
The optimal values of $\beta $ and $\gamma $ are obtained numerically by
minimizing the ground-state energy according to $\partial E_{gs}^{\mathtt{%
GSRWA}}/\partial \beta =0$ and $\partial E_{gs}^{\mathtt{GSRWA}}/\partial
\lambda =0$. By setting the squeezing $\lambda =0$ in Eq.(~\ref{gs energy}),
the displacement $\beta $ in the GVM is optimized by minimizing the
ground-state energy
\begin{equation}
E_{gs}^{\mathtt{GVM}}=\omega \beta ^{2}-2\beta \alpha -(\frac{\Delta }{2}%
+2\gamma \beta )e^{-2\beta ^{2}}.
\end{equation}%
Meanwhile, the ground-state energy in the GRWA with $\beta =\alpha /\omega $
is obtained easily as $E_{gs}^{\mathtt{GRWA}}=-\alpha ^{2}/\omega -(\frac{%
\Delta }{2}+2\gamma \alpha /\omega )e^{-2\alpha ^{2}/\omega ^{2}}$.

Figure~\ref{large g2} displays the ground-state energy $E_{gs}/\omega$ as a
function of the coupling strength $g_1$ for $g_2=1.5g_1$ under different
detunings $\Delta/\omega=2$ (a) and $\Delta/\omega=1$ (b) by means of four
methods. For the coupling strength $g_1/\omega<0.5$, both results of the GSRWA and
the GVM agree well with the numerical ones. As $g_1/\omega$ increases to $0.8$,
the GVM deviates from the numerical results. This
deviation becomes more obvious for positive detuning $\Delta/\omega=2$.
However, the GSRWA results coincide with those of the numerical simulation
well in the ultra-strong coupling regime except for a small discrepancy in the
deep-strong coupling regime for $\Delta/\omega=2$. The GRWA produces
unreasonable results in the ultra-strong coupling regime. It is
demonstrated that the GSRWA results are much better than the GVM and GRWA ones
in the anisotropic case.

The advantage of the GSRWA lies in the contribution from the variable
squeezing and displacement. The optimum values for $\beta$ and $\lambda$
have interesting dependences on the coupling $g_1/\omega$ and the detunings $%
\Delta/\omega$, as shown in Figs.~\ref{large g2}(c) and ~\ref{large g2}(d). As $g_1/\omega$
increases, effects of both displacement and squeezing transformations become
to play an important role in lowering the ground-state energy. $\beta$ and $%
\lambda$ develop, and in particular, the squeezing $\lambda$ increases as
the coupling strengthens to a maximum value, and then decreases in the
deep-strong coupling regime, which is qualitatively similar to the
prediction by the analytical $\lambda$ in the isotropic case. And the larger $%
\Delta/\omega$ is chosen, the larger the value of the maximum $\lambda_{max}$
reaches. Thus the squeezing effects can not be underestimated
with the increase in $\Delta/\omega$. In the deep-strong coupling limit, $%
\beta$ nearly increases linearly with $g_1/\omega$, and $\lambda$ decays to
zero, which is the result predicted by the GRWA. Therefore, the contributions of
both the displacement and squeezing transformations are significant
especially for the ultra-strong coupling strength $g_1/\omega\sim1$, where the GVM
and the GRWA are no longer valid.

\begin{figure}[tbp]
\includegraphics[scale=0.7]{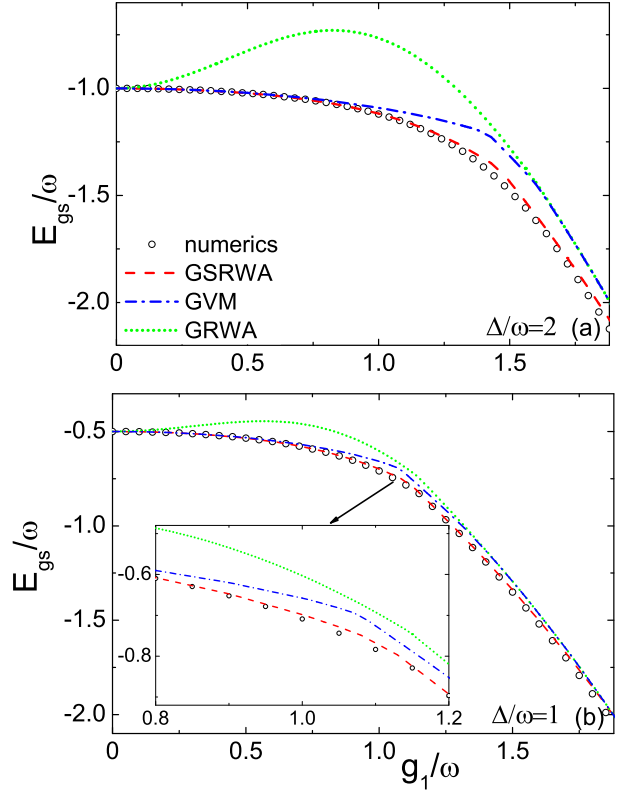}
\caption{(Color online) Ground-state energy $E_{gs}/\protect\omega$ in the
anisotropic Rabi model with $\protect\gamma<0$ as a function of $g_1/\protect%
\omega$ by means of the GSRWA (dashed lines), numerical simulation
(circles), GVM (dashed dotted lines) and GRWA (dotted lines) for $g_2=0.5g_1$
for different detunings (a)$\Delta/\protect\omega=2$ and (b) $\Delta/\protect%
\omega=1$.}
\label{weak g2}
\end{figure}

\textsl{Dominated coupling of the RW interaction ($\gamma <0$):} For the
anisotropic case of the dominated coupling of the RW interaction with $%
\gamma <0$, $g_{2}<g_{1}$, energy-level crossing between the ground state
and the first-excited state occurs due to the competition between the
rotating and counter-rotating interactions~\cite{xie}. And the first excited
state consisting of $|0\rangle |+z\rangle $ and $|1\rangle |-z\rangle $ is
expected to be the ground state when the coupling strength exceeds the
crossing point~\cite{xie}.

In terms of the basis $|0\rangle |+z\rangle $ and $|1\rangle |-z\rangle $,
the transformed Hamiltonian (~\ref{effective ham}) obtained by the GSRWA is
written in the matrix form as
\begin{equation}
H_{\mathtt{GSRWA}}=\left(
\begin{array}{cc}
\eta _{0}+f(0) & \eta _{2}(\alpha -\beta )+P_{1,0} \\
\eta _{2}(\alpha -\beta )+P_{1,0} & \eta _{0}+\eta _{1}-f(1)%
\end{array}%
\right) ,
\end{equation}%
$\allowbreak $where $P_{1,0}=\frac{\Delta }{2}R_{1,0}-\gamma \eta T_{1,0}$.
The eigenvalue of the first exited state is given by
\begin{eqnarray}
E_{gs}^{\mathtt{GSRWA}} &=&\eta _{0}+\frac{\eta _{1}+f(0)-f(1)}{2}  \notag \\
&&-\frac{1}{2}\sqrt{[f(0)-\eta _{1}+f(1)]^{2}+4P_{1,0}^{2}},
\label{first energy}
\end{eqnarray}%
and the corresponding first-excited state takes the form as
\begin{equation}
|\varphi _{1}\rangle =\cos \theta |0,+z\rangle +\sin \theta |1,-z\rangle ,
\end{equation}%
with $\tan \theta =\frac{2P_{1,0}}{f(1)+f(0)-\eta _{1}-\sqrt{[f(0)-\eta
_{1}+f(1)]^{2}+4P_{1,0}^{2}}}$.

For the crossing of the ground state and the first-excited state, the
crossing point $g_c$ is determined numerically when the ground-state energy
(~\ref{gs energy}) is equivalent to the first-excited energy, $E_{1}^{\mathtt{%
GSRWA}}=E_{gs}^{\mathtt{GSRWA}}$. After the crossing point, $\beta $ and $%
\lambda $ are determined variationally according to the equations of
minimizing the first-excited energy by $\partial E_{1}^{\mathtt{GSRWA}%
}/\partial \beta =0$ and $\partial E_{1}^{\mathtt{GSRWA}}/\partial \lambda
=0 $. Similarly, one can obtain the corresponding ground-state energy after
the crossing point by setting $\lambda =0$ for the GVM and $\beta =\alpha
/\omega $, $\lambda =0$ for the GRWA in Eq. (~\ref{first energy}).

Figure~\ref{weak g2} shows the ground-state energy $E_{gs}$ as a function of
the coupling strength $g_1/\omega$ by different methods under different
detunings. $E_{gs}$ obtained by the GSRWA is consistent with the numerical
ones even for the ultra-strong coupling strength up to $g_1/\omega\sim1$. However, we can see that the GVM with
only the variable displacement deviates from the numerical results in the
ultra-strong coupling regime $g_1/\omega>0.8$, and the one by the fixed $\beta$
under the GRWA shows a dramatic deviation even in the weak coupling regime.
Moreover, the GVM and the GRWA get worse as the detuning $\Delta/\omega$
increases to $2$ in Fig.~\ref{weak g2}(a). It exhibits an overall
improvement of the GSRWA by adding a dimensionless variable squeezing to the
original GVM or GRWA in the anisotropic Rabi model.

\section{Conclusion}

We present the generalized squeezing rotating-wave approximation,
which depends on the optimal displacement and squeezing
parameters, as determined variationally. The approach by adding the squeezing
transformation improves the previous generalized variational method and generalized RWA only with the displacement
transformation. The ground state is obtained as a well-defined Schr\"{o}%
dinger-cat-like entangled state with the displaced-squeezed oscillator state
instead of the displaced state.
For the isotropic Rabi case, the analytical expression of the ground-state energy and the mean
photon number agree well with the numerical ones especially for the
ultra-strong coupling strength up to $g_1/\omega\sim1$, whereas the previous
results with only the displacement show distinguished
deviation for large atom frequency. For the anisotropic case, the
ground-state energy obtained by the presented approach is more accurate than
previous results. It demonstrates that the contribution of the squeezing is significant
in the ground state in the ultra-strong coupling regime especially for large atom frequency.
In additional, an efficient treatment for excited states is expected to be proposed
in future work by considering the squeezing effect. The
analytical solution presented here can easily be implemented to simulate
superconducting qubit-oscillator systems for ultrastrong-coupling strengths up to
$g/\omega\sim1$, which is accessible in recent circuit QED experiment setups~%
\cite{yoshihara}.
\section{Acknowledgments}

This work was supported by the Chongqing Research Program of Basic Research and
Frontier Technology (Grant No.cstc2015jcyjA00043), and the Research Fund for the
Central Universities (Grants No.106112016CDJXY300005, and No. CQDXWL-2014-Z006).

Email:yuyuzh@cqu.edu.cn

\end{document}